# Exploring atmospheric radon with airborne gamma-ray spectroscopy

Marica Baldoncini[1,2], Matteo Albéri[1,2], Carlo Bottardi[1], Brian Minty[3], Kassandra G.C. Raptis[1,4], Virginia Strati[1,4], and Fabio Mantovani[1,2]

[1]Department of Physics and Earth Sciences, University of Ferrara, Via Saragat 1, 44121 - Ferrara, Italy
[2]INFN, Ferrara Section, Via Saragat 1, 44121 - Ferrara, Italy
[3]Minty Geophysics, PO Box 3299, Weston Creek, ACT, 2611, Australia
[4]INFN, Legnaro National Laboratories, Viale dell'Universitá, 2 - 35020 Legnaro (Padua), Italy

*Correspondence to:* Marica Baldoncini (baldoncini@fe.infn.it)

**Abstract.**
$^{222}$Rn is a noble radioactive gas produced along the $^{238}$U decay chain, which is present in the majority of soils and rocks. As $^{222}$Rn is the most relevant source of natural background radiation, understanding its distribution in the environment is of great concern for investigating the health impacts of low-level radioactivity and for supporting regulation of human exposure to ionizing radiation in modern society. At the same time, $^{222}$Rn is a widespread atmospheric tracer whose spatial distribution is generally used as a proxy for climate and pollution studies. Airborne gamma-ray spectroscopy (AGRS) always treated $^{222}$Rn as a source of background since it affects the indirect estimate of equivalent $^{238}$U concentration. In this work the AGRS method is used for the first time for quantifying the presence of $^{222}$Rn in the atmosphere and assessing its vertical profile. High statistics radiometric data acquired during an offshore survey are fitted as a superposition of a constant component due to the experimental setup background radioactivity plus a height dependent contribution due to cosmic radiation and atmospheric $^{222}$Rn. The refined statistical analysis provides not only a conclusive evidence of AGRS $^{222}$Rn detection but also a (0.96 ± 0.07) Bq/m$^3$ $^{222}$Rn concentration and a (1318 ± 22) m atmospheric layer depth fully compatible with literature data.

## 1 Introduction

$^{222}$Rn is a naturally occurring noble gas produced via alpha decay of $^{226}$Ra and it is the only gaseous daughter product of the decay chain of $^{238}$U, which is present in the majority of soil and rock types and which has a half-life of ~4.5·10$^9$ yr, comparable to the Earth's age. As $^{222}$Rn is almost chemically inert, it exhales from soils and rocks into the atmosphere and migrates by diffusion and convection without being subject to atmospheric removal processes, therefore running out mainly through radioactive decay (*Jacobi and André*, 1963). $^{222}$Rn atmospheric abundance is strictly connected with its exhalation rate from soils, which is typically on the order of 0.2 - 1.5 atoms/(cm$^2$·s) (*Beck*, 1974) which is in turn affected by soil type,

granulometry and moisture content, as well as by porosity and permeability (*Turekian et al.*, 1977; *Szegvary et al.*, 2009; *Manohar et al.*, 2013).

Radon gas is responsible for the largest human exposure to natural ionizing radiation, the majority of which takes place in the home (*UNSCEAR*, 2008; *WHO*, 2009): in this context, the characterization of building materials and drinking water is considered a relevant topic in the field of radiation protection (*Chen et al.*, 2010; *Hulka*, 2008; *Nuccetelli and Bolzan*, 2001; *Rizzo et al.*, 2001; *Messier et al.*, 2015). In the light of assessing human exposure to radon radiation, strong efforts are being devoted to combine information coming from indoor radon measurements, airborne gamma-ray (AGRS) spectroscopy measurements and geological mapping (*Smethurst et al.*, 2017; *Appleton et al.*, 2011).

The poor chemical reactivity, together with the 3.82 days half-life, makes $^{222}$Rn a conventional and widespread atmospheric tracer. Indeed, $^{222}$Rn has a relatively long half-life for being connotative of events related to turbulence (having a typical 1 hour time scale), but it also lasts shortly enough to have a high concentration gradient through the lower troposphere that can give insights into air vertical mixing mechanisms and help in tracing air transport processes. Monitoring atmospheric $^{222}$Rn has a variety of applications in climate, air quality and pollution studies, including tracing air mass transport, tracing diurnal mixing in the lower atmosphere, calibrating seasonal regional emissions of climatically sensitive tracers including $CO_2$, $CH_4$, $N_2O$, and validating transport and mixing schemes in climate/weather models (*IAEA*, 2012). In the past a great effort has been dedicated in modeling the radon flux and air transport in the atmospheric boundary layer over land disregarding the contribution coming from the ocean, but recently it has been found that, although radon flux density from the ocean is typically few tens percent compared with average flux density from the land, it can provide significant contributions for specific wind conditions (*Schery and Huang*, 2004).

Measurements of the vertical distribution of $^{222}$Rn can be conducted as tower-based studies, which generally have high vertical resolution but altitude limited to 5 - 40 m, as well as via airborne $^{222}$Rn or $^{222}$Rn progeny measurements, which can span a larger height range (from hundreds of m to more than 10 km) but typically resolve few altitudes (*Williams et al.*, 2010). Direct $^{222}$Rn measurements are generally carried out by filling scintillation Lucas cells with laboratory extracted $^{222}$Rn absorbed onto activated charcoal after exposure to sampled air, while indirect measurements are generally made by alpha counting of $^{222}$Rn progeny (*Baskaran*, 2016). The former provides direct radon concentrations, even if having an extracting and counting apparatus at short distance is necessary in order to reduce the time available for $^{222}$Rn to decay. On the other hand, $^{222}$Rn progeny measurements rely on the assumption of secular equilibrium between $^{222}$Rn and its daughter products.

Variations in the vertical radon concentration profiles produce changes in the natural background gamma-ray flux which, in turn, can be responsible for perturbations and contaminations in aerial monitoring results (*Beck*, 1974). $^{214}$Pb, having a half-life of 26.8 minutes, and $^{214}$Bi, having a half-life of 19.8 minutes, are the two principal gamma-emitting daughters of $^{222}$Rn, which, thanks to their short decay time, are usually in equilibrium with each other (i.e., their activities are about the same at all elevations). When the vertical mixing conditions are not characterized by quick variations (as happens close to sunrise and sunset), the steady state is generally reached which means that the concentration profiles of radon and its daughters tend to be near secular equilibrium, except near ground (h < 25 m) (*Gogolak*, 1977).

In this work we present the results of a ∼4 hours AGRS survey over the sea: when flying offshore no geological gamma signal is detected and the measured spectra result from the superposition of a constant contribution coming from the radioactivity of the equipment and of the height dependent contributions associated with cosmic radiation and with atmospheric radon. The AGRS campaign has been conducted over a wide range of altitudes, from 77 m up to 3066 m. Thanks to this large elevation extent, it has been possible to explore the presence of radon in the atmosphere via the modeling of the expected count rate in the $^{214}$Bi photopeak energy window according to two analytical models which respectively exclude and account for the presence of atmospheric radon.

## 2 Materials and Methods

### 2.1 Experimental setup, survey and data

Three AGRS surveys have been performed over the Tyrrhenian Sea in proximity of Viareggio (Tuscany, Italy) with a prototype autogyro called Radgyro (see Fig. 1), whose engineering has been expressly devised in order to make the aircraft a flying multisensorial platform devoted to measurements of electromagnetic waves in the field of proximal remote sensing (*Albéri et al.*, 2017). The Radgyro positioning is carried out by processing with the goGPS software (*Herrera et al.*, 2016) the binary acquisitions of two u-blox EVK-6T GPS antennas for the extraction of the geographic latitude and longitude, together with the orthometric altitude at 1Hz. Gamma-ray measurements are performed with a modular NaI(Tl) scintillation detector arranged in the middle of the Radgyro hull, the AGRS_16L, which is made up of 4 4L crystals having dimensions equal to 10 cm × 10 cm × 40 cm (*Guastaldi et al.*, 2013; *Strati et al.*, 2015). The acquired list mode files, reporting for each ADC channel the energy deposition inside the specific crystal, are processed offline in order to generate for each detector 1Hz energy calibrated gamma spectra, summed up to obtain the gamma-ray spectrum resulting from the whole 16L detection volume (*Baldoncini et al.*, 2017). Radiometric data entering this analysis have been collected at a maximum distance from the coast of about 4.5 km and have been selected by requiring a minimum distance of 300 m, which is meant to exclude gamma-ray signals potentially spoiled by ground radiation. According to this selection cut, the overall effective acquisition statistics for the three flights is 14688 seconds, as reported in Table 1 along with the main features referred to the single surveys.

The estimated count rates in the energy windows of interest have been clustered in altitude bins of 15 m, which is conservative with respect to the estimated accuracy of the vertical position determined by the instrumental setup (*Albéri et al.*, 2017). The count rates are estimated by summing all the input count rates acquired in the same elevation bin and dividing by the number of 1 second spectra entering the summation. Figure 2 shows the count rates measured respectively in the $^{214}$Bi Energy Window (BEW, 1.66 - 1.86 MeV), $^{208}$Tl Energy Window (TEW, 2.41 - 2.81 MeV) and Cosmic Energy Window (CEW, 3.0 - 7.0 MeV) as function of the altitude above sea level, distinguished according to the different flights. In the TEW and CEW, separately, the variation of the count rates in different flights is compatible with the statistical fluctuation of the count rates: there is no systematic effect related to the different flight times and the exponential behavior is maintained down to low elevations. For the count rates in the BEW there is some evidence of data clustering for different flights, in particular at low elevations, which is a hint of the presence of $^{222}$Rn gas in the atmosphere.

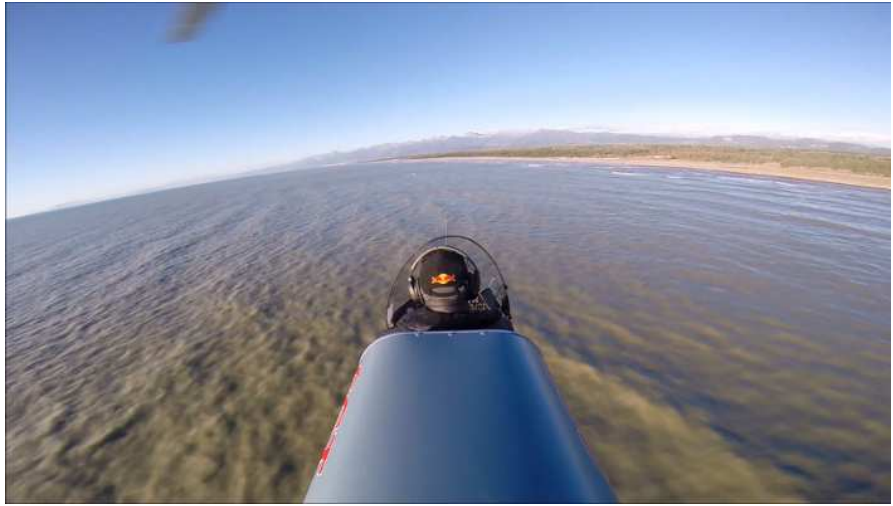

**Figure 1.** Picture of the Radgyro taken during the airborne gamma-ray survey over the sea.

**Table 1.** Summary of the main parameters for each of the 3 surveys over the sea. In the case of flights 11 and 14, 83 seconds and 30 seconds have been cut due to some radiofrequency interference between the PMT and the aircraft transponder. For each flight we report the ID, date, time, minimum and maximum altitude and acquisition time together with ground temperature (T), ground pressure (P) and ground wind velocity (W) at the take off and at the landing and sky conditions. Globally the weather conditions during the flights were stable and without precipitations.

| Flight ID | Date | Time | z min [m] | z max [m] | Acquisition time | T [°C] | P [hPa] | W [km/h] | Sky conditions |
|---|---|---|---|---|---|---|---|---|---|
| 11 | 30/03/2016 | 17:42:10 | 77 | 2019 | 6370 | 18.6 | 1016.8 | 17 | Mostly clear |
|  |  | 19:29:43 |  |  |  | 14.9 | 1015.3 | 15 |  |
| 12 | 31/03/2016 | 18:13:55 | 126 | 2070 | 3041 | 22.2 | 1010.3 | 17 | Mostly clear |
|  |  | 19:46:47 |  |  |  | 19.7 | 1009.9 | 11 |  |
| 14 | 05/04/2016 | 16:37:16 | 461 | 3066 | 5277 | 24.6 | 1007.2 | 17 | Clear |
|  |  | 18:05:43 |  |  |  | 20.7 | 1015.7 | 2 |  |
| Global |  |  | 77 | 3066 | 14688 |  |  |  |  |

## 2.2 Theoretical Model

$^{222}$Rn daughter products $^{214}$Pb and $^{214}$Bi are the main gamma-emitters in the $^{238}$U decay chain and, since they bind to airborne aerosols, they are responsible for the measured radon background. Estimates of the $^{238}$U content via AGRS measurements rely on the evaluation of background subtracted count rates in the $^{214}$Bi photopeak energy window (BEW), which corresponds to the (1660-1860) keV energy range centered on the 1765 keV $^{214}$Bi gamma emission line. Background correction involves the removal of gamma signal of non-geologic nature, which consists of three components resulting respectively from the decay of $^{214}$Bi in the atmosphere, the radioactivity of the aircraft and its equipment due to presence of trace amounts

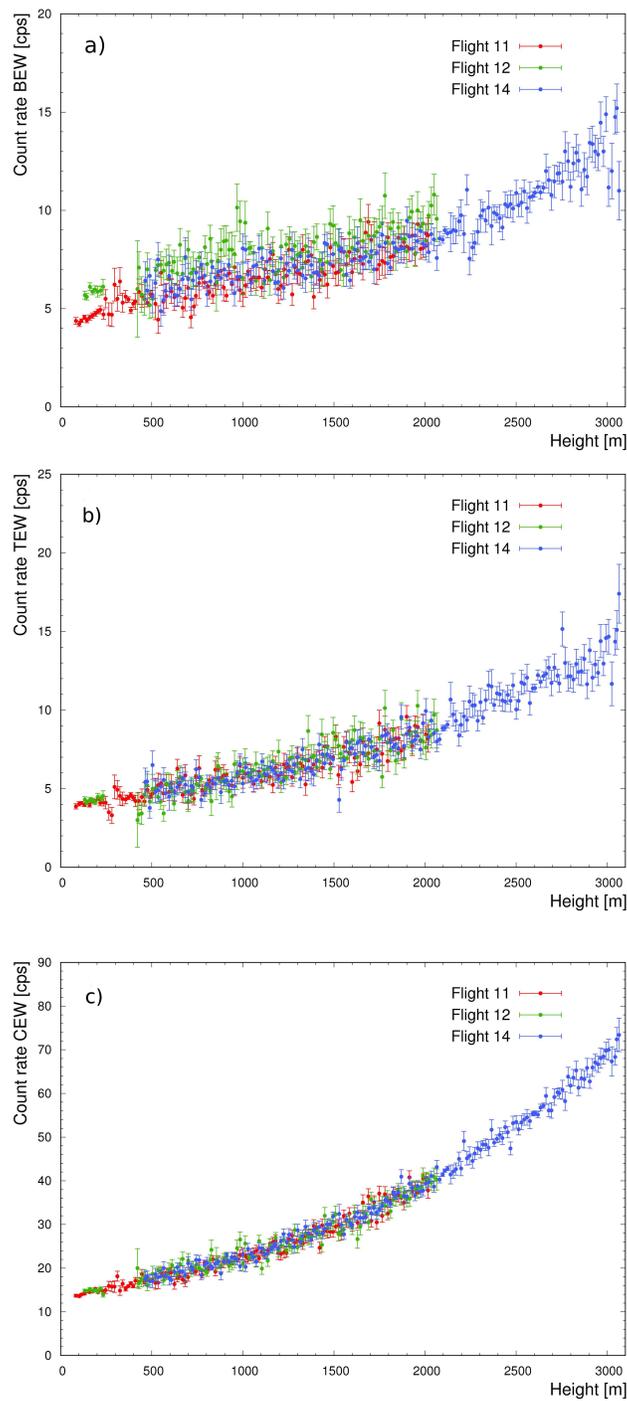

**Figure 2.** Panels a), b) and c) show the count rate respectively in the BEW, TEW and CEW as function of the altitude for the 3 different flights carried out during the survey over the sea. Both in the TEW and in the CEW experimental data from different flights sit on top of each other, excluding systematic effects associated to the different acquisition times. In the BEW it is possible to recognize the effect of atmospheric radon contamination for the 3 different flights.

of $^{238}$U and $^{232}$Th, and the interaction of secondary cosmic radiation with the air, the aircraft and the detector (*Minty*, 1998). AGRS detectors are generally calibrated for the aircraft and cosmic background by performing high-altitude offshore flights in an area where atmospheric radon is at minimum. *IAEA* (2003) suggests to measure spectra at a range of heights, typically from 1.0 - 1.5 km up to 3.0 - 3.5 km over water with a 300 - 500 m step, for generally 10 - 15 minutes accumulation time at each height. In the absence of radon gas, the count rate in the BEW can be described as a superposition of a constant aircraft component and a cosmic component which is expected to exponentially increase with increasing height above sea level as stated by the following equation:

$$n_{BEW}^{aircraft+cosmic}(z) = A_{BEW} e^{\mu_{BEW} z} + B_{BEW} \qquad (1)$$

where $n_{BEW}^{aircraft+cosmic}(z)$ is the count rate in the BEW and $A_{BEW}$, $\mu_{BEW}$ and $B_{BEW}$ are constants (*Grasty and Minty*, 1995; *IAEA*, 1991). This radon free model is expected to accommodate experimental measurements, generally at altitudes greater than 2000 m. Indeed, although the atmospheric concentration of $^{222}$Rn and of its daughter products can vary significantly with different diffusion conditions, mean $^{222}$Rn concentrations are (4 $\pm$ 3) Bq/m$^3$ in the lowest 30 - 1000 m, while above 1000 - 1500 m mean $^{222}$Rn concentrations generally show a steep decrease to values compatible with zero (around (2 $\pm$ 2) Bq/m$^3$), dropping even further to (0.3 $\pm$ 0.4) Bq/m$^3$ above 3000 m (*Williams et al.*, 2010). When looking to experimental data acquired at low altitudes, a deviation from the mentioned exponential behavior can be observed due to radon accumulation in the atmosphere. Traditionally, the presence of atmospheric radon is identified as a breakdown of the linear relation that is supposed to hold between the count rates in the BEW and the count rates measured in the CEW, the latter having exclusively cosmic origin since the maximum terrestrial gamma energy corresponds to the 2614 keV $^{208}$Tl emission (*Grasty and Minty*, 1995).

An alternative model can be developed with the aim of covering the entire altitude range and of recognizing and possibly quantifying the presence of the radon gas in the atmosphere via the detection of the gamma-signal generated by the $^{214}$Bi decay. In presence of atmospheric radon, the overall count rate recorded in the BEW $n_{BEW}(z)$ comprises not only the aircraft plus cosmic component $n_{BEW}^{aircraft+cosmic}(z)$ (see Eq. 1) but also an altitude dependent component arising from atmospheric $^{214}$Bi ($n_{BEW}^{Rn}(z)$) whose modeling requires a radon vertical profile, which is in turn directly connected with the dynamics of the atmospheric boundary layer.

The diurnal evolution of the atmospheric boundary layer, i.e. the ∼1-2 km thick layer where the atmosphere feels the contact with the ground surface, is governed by the mechanical and thermal surface-air interactions which are respectively driven by wind and solar radiation. Under clear sky conditions, after sunrise the warmed ground heats the air touching the ground, creating thermals that rise and cause intense motions which gradually create a convective boundary layer (or mixed layer), generally characterized by high homogeneity. As time passes, the growing convective region reaches higher altitudes till at sunset thermals cease and convection terminates, leading to the formation of a residual layer containing near zero turbulence and the residual moisture, heat, and pollutants that were mixed during the day. As long as the weather remains fair the cycle repeats on a daily timescale, with a mixing efficiency that partially depends on the amount of cover due to clouds which can intercept portions of the sunlight and reduce the amount of heat delivered to ground level (*Stull*, 2012).

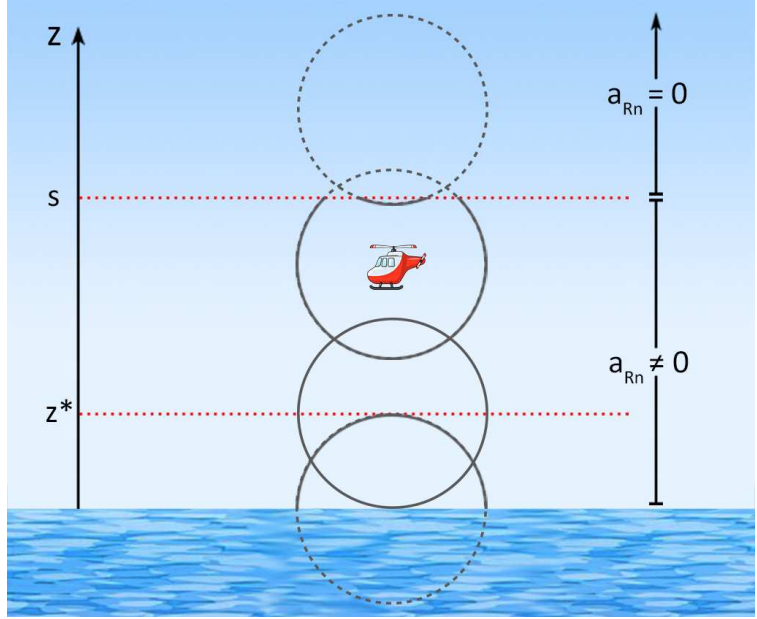

**Figure 3.** Schematic illustration of the variation of the detector field of view to the atmospheric $^{214}$Bi gamma signal with respect to the height. When the detector is at sea level, the field of view will be that of a half-sphere. With increasing height the detector starts seeing the upward photon flux till the field of view reaches saturation at the altitude z*, corresponding to the full-sphere case. Approaching the separation altitude s between the two radon layers the field of view starts shrinking and finally vanishes when the detector is completely immersed in the radon free layer.

In cases of fair weather, for convective boundary layers a very marked drop in radon concentrations is generally observed in crossing the separation between the mixed layer and the free troposphere, where radon abundances reach typically near-zero values (*Williams et al.*, 2010). In the case of mixed layers topped with residual layers radon exhibits a fairly constant profile in the mixed layer and tends to reduce linearly with height in the residual layers.

As the airborne campaign was conducted under clear sky conditions in a narrow range of days and always in the late afternoon, the simplified radon vertical profile adopted in this study is a discrete model according to which the radon concentration is uniform up to a cutoff altitude $s$, basically corresponding to the depth of the mixed layer, and null above the cutoff height. Figure 3 shows a schematic example of the behavior of the field of view of the gamma-ray detector to $^{214}$Bi gamma signal as it moves to increasing altitude, starting from sea level up to the separation height between the two radon gas layers, till it reaches the radon free zone.

In the lower layer where the radon activity is uniform, the contribution to the count rate in the BEW originated by the atmospheric $^{214}$Bi has a monotonic increase with increasing altitude. Indeed, at altitude zero the detector field of view can be approximated by a half-sphere as the gamma photon flux has only a downward incoming direction; when the detector starts lifting from sea level an upward incoming photon flux will start being visible enhancing the detected gamma signal. At an altitude equal to half the separation height $n_{BEW}^{Rn}(z)$ will reach its maximum. If the cutoff altitude $s$ is high enough (for $s >$

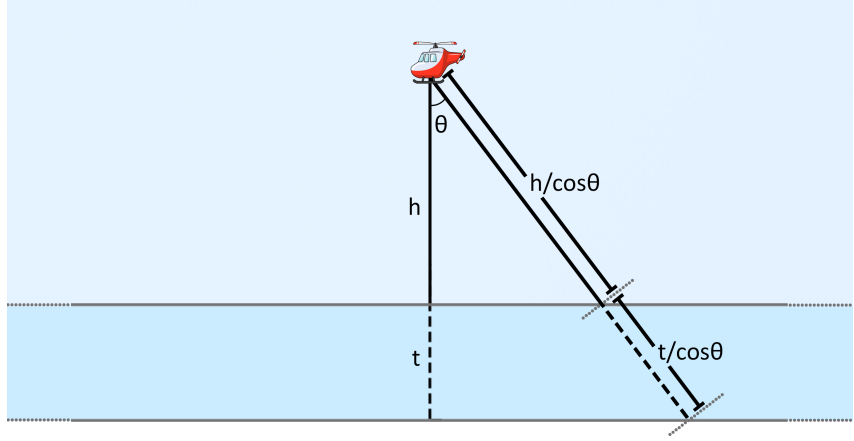

**Figure 4.** Schematic diagram of the geometrical model adopted for estimating the unscattered photon flux reaching a detector situated at a vertical distance h from a source having infinite lateral extension and thickness t. In this context the source of thickness t corresponds to an air layer in which a homogeneous radon concentration is present.

400 m, corresponding to $\sim$2.3 photon mean free paths, the count rate is essentially constant), the maximum count rate will reach a saturation value almost equal to double the count rate recorded at sea level, corresponding to the full-sphere field of view. Approaching the separation height $s$, the $n_{BEW}^{Rn}(z)$ count rate will start monotonically decreasing till it vanishes when the detector is far enough from the lower radon layer.

From the theoretical point of view it is necessary to model the propagation of unscattered photons from the source to the detector position (Figure 4). By integrating in spherical coordinates and by taking into account the azimuthal symmetry of the model, the flux of unscattered 1765 keV photons emitted by atmospheric $^{214}$Bi is given by the following equation:

$$\Phi = \frac{A_v P_\gamma}{2\mu_a} \int_0^1 dcos\theta e^{\frac{-\mu_a h}{cos\theta}} \left[ 1 - e^{\frac{-\mu_a t}{cos\theta}} \right] \tag{2}$$

where $A_v$ is the volumetric activity in [Bq/m$^3$] of the uniformly distributed $^{214}$Bi, $P_\gamma$ is the $\gamma$-ray intensity for 1765 keV photons in [# of emitted $\gamma$/Bq], $\mu_a$ is the air linear attenuation coefficient referred to 1765 keV photons, $t$ is the thickness of the air layer in which gamma photons are homogeneously and isotropically emitted, $h$ is the vertical distance of the detector from the source layer (*Feng et al.*, 2009). By scaling for the detector cross sectional area and by some efficiency factor, Eq. 2 directly translates into the expression describing the variation of the count rate as a function of altitude.

The $n_{BEW}^{Rn}(z)$ vertical profile can be modeled by distinguishing the case in which the detector vertical position $z$ is below or above the cutoff altitude $s$. In both scenarios the air layer at an altitude greater than $s$ does not give any contribution to the signal as it has zero activity volume concentration. As illustrated in Fig. 5a, when the detector position $z$ is below the cutoff altitude $s$, two air source layers having thickness respectively equal to $z$ and $s-z$ contribute to the radon count rate with $n_1(z)$ and $n_2(z)$ as stated by the following equation:

$$n_{BEW}^{Rn}(z) = n_1(z) + n_2(z) = C \int_0^1 dcos\theta \left[1 - e^{\frac{-\mu_a z}{cos\theta}}\right] + C \int_0^1 dcos\theta \left[1 - e^{\frac{-\mu_a(s-z)}{cos\theta}}\right] \quad (z < s) \qquad (3)$$

where $C$ is the count rate in cps measured at zero distance from a semi-infinite homogeneous air volume source, i.e. the count rate obtained for $h = 0$ and $t \to \infty$ (see Eq. 2). If the detector position is above the cutoff altitude ($z > s$), the count rate arises only from layer number 3 (see Fig. 5b), where the air source layer thickness is $s$ and the detector vertical distance from the source is $z - s$, corresponding to:

$$n_{BEW}^{Rn}(z) = n_3(z) = C \int_0^1 dcos\theta e^{\frac{-\mu_a(z-s)}{cos\theta}} \left[1 - e^{\frac{-\mu_a s}{cos\theta}}\right] \qquad (4)$$

Therefore, the theoretical expression for the count rate in the BEW $n_{BEW}^{Rn}(z)$ can be summarized according to the following equation:

$$n_{BEW}^{Rn}(z) = \Theta(s-z)\left[n_1(z) + n_2(z)\right] + \Theta(z-s)n_3(z) \qquad (5)$$

where $\Theta(x)$ represents the Heaviside step function.

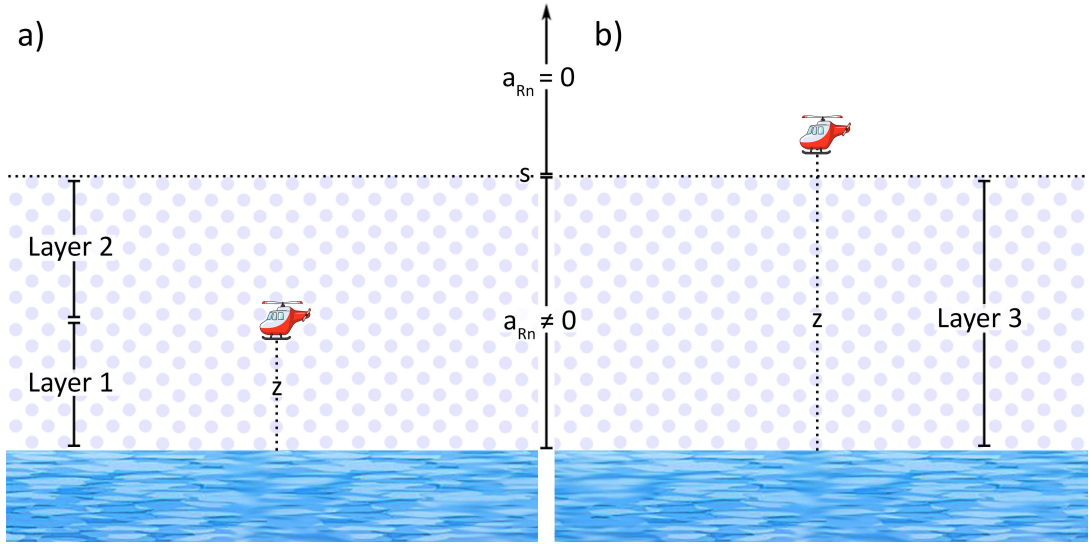

**Figure 5.** Schematic illustration of the air layers generating the radon contribution to the count rate in the BEW. When the detector vertical position $z$ is below the cutoff altitude $s$ (which separates the lower atmospheric portion having uniform radon concentration from the upper one which has null radon abundance), there are two layers generating the $^{214}$Bi gamma signal (a). When the detector vertical position $z$ is above the cutoff altitude $s$, there is only one layer generating the $^{214}$Bi gamma signal (b).

Figure 6 shows a representative example of the $n_{BEW}^{Rn}(z)$ curve. As expected, the curve is symmetrical with respect to an altitude value equal to half the separation height $s$. The separation altitude $s$ corresponds to ~8.7 photon mean free paths, which is a long enough distance for the count rate at sea level $n_{BEW}^{Rn}(0)$ to reach the $C$ value, corresponding to the count rate associated to a semi-infinite volume source. Similarly, $n_{BEW}^{Rn}(z)$ gets to reach and maintain the saturation value equal to $2C$ before starting to decrease when the altitude approaches $s$.

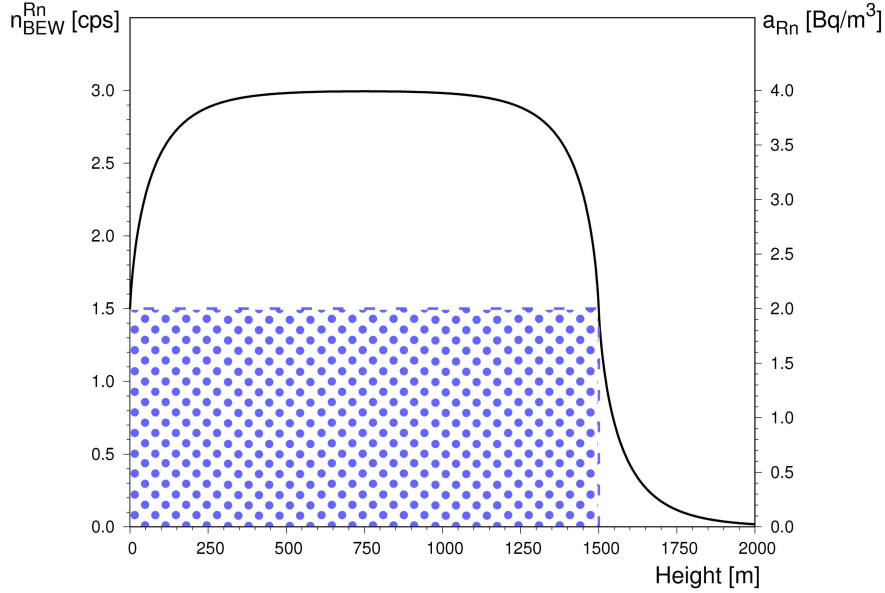

**Figure 6.** The black solid line illustrates the $n_{BEW}^{Rn}(z)$ count rate (left y axis) as function of the altitude for a $C$ count rate value equal to 1.5 cps, a gamma linear attenuation coefficient $\mu_a$ equal to 0.005829 m$^{-1}$ and a cutoff altitude $s$ equal to 1500 m (see Eq. 5). The blue polka-dotted pattern represents the 2 Bq/m$^3$ homogeneous radon concentration (right y axis) in the atmospheric layer below 1500 m. In the air layer at altitude larger than 1500 m the radon concentration vanishes.

The overall count rate in the BEW can be therefore expressed according to the following equation:

$$n_{BEW}(z) = A_{BEW}e^{\mu_{BEW}z} + B_{BEW} + \Theta(s-z)\left[n_1(z) + n_2(z)\right] + \Theta(z-s)n_3(z) \tag{6}$$

Figure 7 shows the global behavior of $n_{BEW}(z)$, together with the separate components associated with the aircraft plus cosmic background and with the radon background. The radon contribution produces a curvature in the model function which is evident in the low altitude range ($z < 200$ m) where the initial half-spherical field of view approaches a full-spherical field of view. After the radon component has reached the plateau, the model curve grows in parallel to the radon free curve just shifted upward by the radon saturation count rate. In approaching the separation altitude between the two radon layers the model curve exhibits a kink, whose vertical extent depends on the values of the exponential function parameters and of the radon concentration gradient between the two layers. This kink translates into a local count rate decrease till the model curve matches the curve obtained in the radon free scenario at an altitude which is ∼400 m higher than the cutoff altitude.

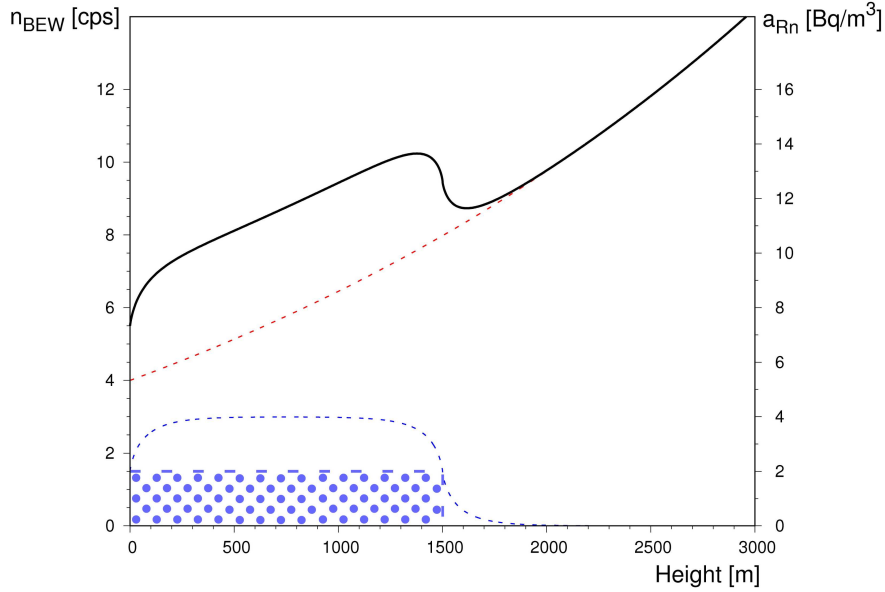

**Figure 7.** The blue dashed line shows the curve for the count rate in the BEW (left y axis) due to the presence of atmospheric radon $n_{BEW}^{Rn}(z)$ obtained for a cutoff altitude $s$ equal to 1500 m and a $C$ value of 1.5 cps (see Eq. 5). The blue polka-dotted pattern represents the 2 Bq/m$^3$ homogeneous radon concentration (right y axis) in the atmospheric layer below 1500 m. In the air layer at altitude larger than 1500 m the radon concentration vanishes. The red dashed line shows the aircraft plus cosmic contribution obtained with $A_{BEW}$ = 7 cps, $\mu_{BEW}$ = 3 $\cdot 10^{-4}$ m$^{-1}$ and $B_{BEW}$ = -3 cps (see Eq. 1). The black solid line represents the overall count rate in the BEW, determined as the sum of the aircraft plus cosmic contribution and the atmospheric radon contribution (see Eq. 6).

## 2.3 Determination of the count rate vertical profile parameters

The two theoretical models described in the previous section (i.e. i) a radon free model defined by Eq. 1 and a 1 layer uniform radon model defined by Eq. 6) have been used in order to reconstruct the observed count rate in the BEW as a function of altitude. The parameters of the theoretical curves have been determined via the minimization of a $\chi^2$ function. For the radon free model the $\chi^2$ minimization has been performed for the count rates measured at elevations greater than 2000 m, where the condition of absence of radon is supposed to hold. On the basis of Eq. 1, the following definition of the $\chi^2$ function has been used:

$$\chi^2 = \sum_{j=1}^{N} \left[ \frac{n_{BEW}^j - (A_{BEW} e^{\mu_{BEW} z_j} + B_{BEW})}{\sigma_{n_{BEW}^j}} \right]^2 \tag{7}$$

where $N$ is 79, equal to the number of experimental data measured at $z_j >$ 2000 m, $n_{BEW}^j$ is the count rate in the BEW measured at $z_j$, $z_j$ is the average elevation obtained for the $j-th$ elevation bin and $\sigma_{n_{BEW}^j}$ is the 1 sigma uncertainty associated to the counting statistics, corresponding to the square root of the total counts recorded at $z_j$ in the BEW divided by the acquisition time. For the model containing the radon contribution, the $\chi^2$ minimization has been performed over the entire altitude range corresponding to the 14688 seconds of data taking. On the basis of Eq. 6, the following definition of the $\chi^2$ function has been used:

$$\chi^2 = \sum_{j=1}^{N} \left[ \frac{n_{BEW}^j - (A_{BEW} e^{\mu_{BEW} z_j} + B_{BEW} + \Theta(s - z_j)[n_1(z_j) + n_2(z_j)] + \Theta(z_j - s) n_3(z_j))}{\sigma_{n_{BEW}^j}} \right]^2 \quad (8)$$

where $N$ is 423, equal to the number of experimental data measured in the entire altitude range, and $n_{BEW}^j$, $z_j$, $\sigma_{n_{BEW}^j}$ defined as previously described. The best fit solutions have been found using a fixed value for the 1765 keV gamma linear attenuation coefficient $\mu_a$ equal to 0.005829 m$^{-1}$ [1].

## 3 Results and discussion

Figure 8a) and Figure 8b) show respectively the fitting curves obtained by minimizing the $\chi^2$ function for the radon free model (see Eq. 7) and for the model allowing for the presence of a uniform radon concentration in the atmosphere up to a cutoff altitude (see Eq. 8). The best fit parameters obtained in both cases are reported in Table 2.

**Table 2.** Fit parameters of the model curves defined by Eq. 1 and by Eq. 6 describing the dependence with the altitude of the count rate in the BEW respectively in the absence or presence of atmospheric radon. The last column reports the value of the reduced $\chi^2$ referred to the entire range of investigated altitudes.

| Theoretical model | $A_{BEW} \pm \delta A_{BEW}$ [cps] | $\mu_{BEW} \pm \delta \mu_{BEW}$ [m$^{-1}$] | $B_{BEW} \pm \delta B_{BEW}$ [cps] | $s \pm \delta s$ [m] | $C \pm \delta C$ [cps] | Reduced $\chi^2$ |
|---|---|---|---|---|---|---|
| without Rn (Eq. 1) | $0.39 \pm 0.07$ | $(1.0 \pm 0.1) \cdot 10^{-3}$ | $5.5 \pm 0.3$ | / | / | 5.0 |
| with Rn (Eq. 6) | $8.2 \pm 0.2$ | $(2.54 \pm 0.06) \cdot 10^{-4}$ | $-4.9 \pm 0.2$ | $1318 \pm 22$ | $0.68 \pm 0.05$ | 2.1 |

From this study it emerges that a theoretical model accounting only for the cosmic and aircraft component is not satisfactory in describing the data distribution, especially at low elevations. Indeed, the model allowing for the presence of radon in the atmosphere provides a better fit to the data, as proved by the reduction of the reduced $\chi^2$ value from 5.0 for the radon free model to 2.1 for the model accounting for radon in the atmosphere.

It is also possible to perform a consistency check of the $A_{BEW}$ and $B_{BEW}$ fit parameters considering that their sum corresponds to the expected count rate at zero altitude in the absence of radon ($n_{BEW}^{aircraft+cosmic}|_{z=0}$). The latter quantity can indeed be obtained also from the parameters of the linear function describing the relation between the count rates in the BEW and the count rates in the CEW, i.e.:

$$n_{BEW}^{aircraft+cosmic}|_{z=0} = a_{BEW} + b_{BEW} \cdot n_{CEW}|_{z=0}, \quad (9)$$
$$\text{with } n_{CEW}(z) = A_{CEW} e^{\mu_{CEW} z} + B_{CEW}$$

---
[1]National Institute of Standard and Technology website, http://physics.nist.gov/PhysRefData/Xcom/html/xcom1.html

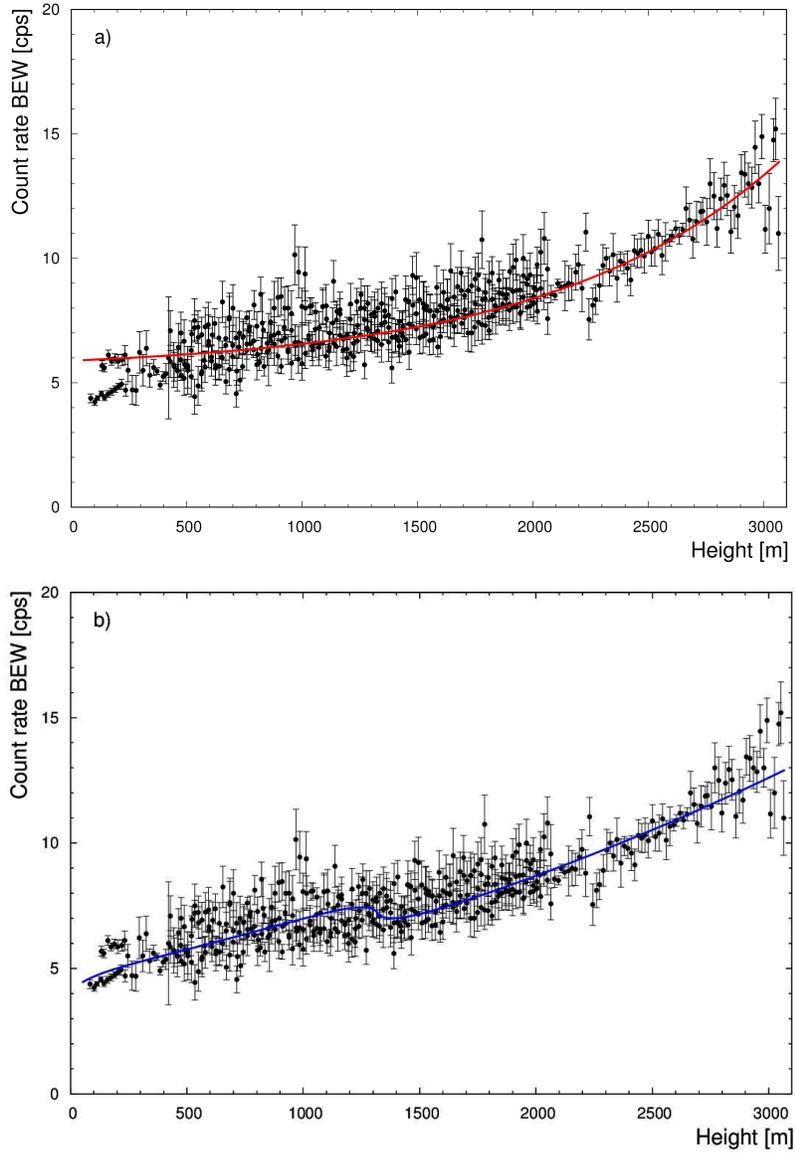

**Figure 8.** Panel a) shows the count rate recorded in the BEW during the entire survey (black points) together with the curve (red solid line) obtained by fitting the data acquired at z>2000 m with a theoretical model that includes only the aircraft and cosmic components of the gamma signal (see Eq. 1). Panel b) shows the same dataset (black points) with the model curve (blue solid line) obtained by fitting the data acquired in the entire elevation range with the theoretical model that accounts also for the presence of radon in the atmosphere (see Eq. 6).

where the fit parameters $a_{BEW}$, $b_{BEW}$, $A_{CEW}$ and $B_{CEW}$ have been obtained in an independent aircraft plus cosmic background calibration survey (*Baldoncini et al.*, 2017)[2].

Therefore, the following equation between fit parameters should hold:

$$A_{BEW} + B_{BEW} = a_{BEW} + b_{BEW}(A_{CEW} + B_{CEW}) \tag{10}$$

The value obtained for the left hand side of Eq. 10 according to the radon free model is (5.9 ± 0.4) cps, while the model accounting for atmospheric radon provides (3.3 ± 0.4) cps, which are respectively incompatible and compatible at $1\sigma$ level with the right hand side value of (4.1 ± 0.7) cps. The fit value for the $s$ parameter is equal to (1318 ± 22) m, comparable with atmospheric radon ranges reported in *Williams et al.* (2010).

The fit value for the $C$ parameter corresponds to the sea level count rate associated to the presence of radon (and its gamma emitting daughter nuclei) in the atmosphere, which can be converted into radon abundance, provided a sensitivity calibration factor. From an independent ground calibration campaign, the sensitivity matrix necessary for the estimation of the natural radionuclide concentrations via the Window Analysis Method has been determined (*IAEA*, 1991). On the base of the calibration process we estimated a sensitivity coefficient $S_{UU} = 0.71$ cps/(Bq/m$^3$). Since it allows for converting the eU volumetric abundance into count rate in the BEW, we can perform a crude estimate of mean radon concentration in the atmospheric (mixed) layer of $Rn$ = (0.96 ± 0.07) Bq/m$^3$. The obtained values for the mean radon abundance and for the mixed layer height are comparable with data published by *Williams et al.* (2010) and *Chen et al.* (2016). In Figure 6 of *Chen et al.* (2016) it is shown that radon concentration is inversely related to the mixing layer height, corresponding typically to about 1 Bq/m$^3$ for a mixing layer height of 1500 m. Moreover, the diurnal variations of radon abundance and mixing layer height in different seasons (Figure 5 of *Chen et al.* (2016)) show that typical values of radon abundance in the spring late afternoon are about 1.2 Bq/m$^3$ for a mixing layer height of ∼1000 m.

A further generalization of the 1 layer model having radon contribution described by Eq. 5 led to the theoretical description of a 2 layers model built by introducing the $s_1$, $s_2$, $C_1$ and $C_2$ model parameters in the mathematical description of the count rate. $s_1$ and $s_2$ correspond to the separation altitudes of a lower and a higher atmospheric layer characterized respectively by a $C_1$ and $C_2$ count rate. The best fit with the 2 layers model provided a $\chi^2 = 2.0$ and $s_1$ = (1166 ± 12) m, $Rn_1$ = (1.24 ± 0.09) Bq/m$^3$, $s_2$ = (1562 ± 28) m and $Rn_2$ = (0.6 ± 0.1) Bq/m$^3$, where $Rn_1$ and $Rn_2$ have been obtained by dividing the $C_1$ and $C_2$ cps values by the $S_{UU}$ constant. The 2 layers and 1 layer models fit the experimental data with essentially the same statistical significance, providing similar $\chi^2$ values. According to the quality of our dataset, it is not feasible to have a clear discrimination between a 1 layer model and a 2 layers model: indeed, the 1 layer model having best fit parameters $s$ = (1318 ± 22) m and $Rn$ = (0.96 ± 0.07) Bq/m$^3$ basically represent the same scenario of a 2 layers model characterized by the above mentioned best fit parameters $s_1$, $s_2$, $C_1$ and $C_2$, where the average separation altitude and the average radon content essentially reproduce the $s$ and $Rn$ values provided by the 1 layer model.

---

[2] $a_{BEW} = (2.0 \pm 0.4)$ cps, $b_{BEW} = (0.16 \pm 0.01)$ [cps in BEW]/[cps in CEW], $A_{CEW} = (11.4 \pm 0.3)$ cps and $B_{CEW} = (2.0 \pm 0.4)$ cps

## 3.1 Conclusions and Perspectives

Radon measurements are typically performed by counting experiments of alpha-particles or beta-particles emitted in the decay of radon progeny, requiring the collection and filtering of air mass samples which is a time consuming and laboratory intense procedure. In this work we proved the feasibility of performing atmospheric $^{214}$Bi AGRS measurements and of assessing its abundance and vertical distribution: in this context, future combined direct radon measurements would provide an important basis for the validation of the model proposed in this paper.

The discrimination of $^{214}$Bi gamma emissions from other sources of radiation is far from trivial: long acquisitions over a wide range of altitudes are a key ingredient for splitting the different contributions to the measured count rates. Indeed, according to the quality of the experimental dataset it has not been possible to statistically discriminate a simplified one layer radon vertical distribution from a more refined two layers radon vertical profile. In perspective, AGRS measurements carried out with large detectors (e.g. the typical 33 L NaI(Tl) systems) mounted on helicopters, which unlike autogyros are able to hover, could provide high statistics experimental data at well separated altitudes potentially increasing the resolution on different $^{222}$Rn vertical strata.

**Author Contributions**

Albéri Matteo, Baldoncini Marica, Mantovani Fabio, Minty Brian and Virginia Strati conceived and designed the work as it is. Albéri Matteo, Baldoncini Marica and Mantovani Fabio took part to the data collection with the Radgyro. The radiometric data analysis and its interpretation have been performed by Baldoncini Marica, Mantovani Fabio and Minty Brian. Figures have been designed by Bottardi Carlo and Raptis Kassandra G. C. Baldoncini Marica drafted the article which is critically revised by Albéri Matteo, Bottardi Carlo, Mantovani Fabio, Minty Brian, Raptis Kassandra G. C. and Strati Virginia. All authors gave final approval of the version to be submitted.

**Acknowledgments**

Funding: This work was partially founded by the National Institute of Nuclear Physics (INFN) through the ITALian RAdioactivity project (ITALRAD) and by the Theoretical Astroparticle Physics (TAsP) research network. The co-authors would like to acknowledge the support of the Geological and Seismic Survey of the Umbria Region (UMBRIARAD), of the University of Ferrara (Fondo di Ateneo per la Ricerca scientifica FAR 2016), of the Project Agroalimentare Intelligente CUP D92I16000030009 and of the MIUR (Ministero dell'Istruzione, dell'Universitá e della Ricerca) under MIUR-PRIN-2012 project.

The authors thank the staff of GeoExplorer Impresa Sociale s.r.l. for their support and Mauro Antongiovanni, Claudio Pagotto, Gianluca Tomasi, Ivan Callegari, Gerti Xhixha, Merita Kaçeli Xhixha and Andrea Motti for their collaboration which made possible the realization of this study. The authors show their gratitude to Enrico Chiarelli, Gianni Fiorentini, Eligio Lisi, Michele Montuschi, Barbara Ricci and Carlos Rossi Alvarez for useful comments and discussions.